\documentclass[runningheads,a4paper]{llncs}
\usepackage{amssymb,amsmath,pifont,footnote}
\usepackage{latexsym,pdfpages}
\usepackage{array}
\usepackage{stmaryrd,ifsym}
\usepackage{algorithmic,algorithm}
\usepackage{cite}
\newtheorem{thm}{Theorem} 

\newtheorem{lem}{Lemma}

\newtheorem{rem}{Remark}

\newtheorem{Obs}{Observation}

\newtheorem{examp}{Example}

 {\begin{list}{}%
         {\setlength{\leftmargin}{#1}}%
         \item[]%
 }
 {\end{list}}

\newcommand{\pfbox}{\hfill\qed}

\newcommand{\Q}{\mathbb{Q}}
\newcommand{\R}{\mathbb{R}}

\newcommand{\Set}[1]{\{ #1 \}}
\newcommand{\Range}[2]{#1 ,\dots, #2}
\newcommand{\RangeSet}[2]{\Set{ \Range{#1}{#2}}}

\newcommand{\Tuple}[1]{\langle #1 \rangle}
\newcommand{\WtFunc}{\mathit{wt}}

\newcommand{\Automat}[1]{\ensuremath{\mathcal{#1}}}

\newcommand{\Heading}[1]{\vspace{-0.25cm}\paragraph{\bf{#1}}}
\newcommand{\BeginProof}{\vspace{-0.25cm}\begin{proof}}

\newcommand{\Comment}[1]{}
\newcommand{\Appendix}[1]{}

\newcommand{\MAX}{\max}
\newcommand{\MIN}{\min}
\newcommand{\SUM}{\operatorname{sum}}
\newcommand{\OP}{\operatorname{op}}

\newcommand{\LimInfAvg}{\mathit{LimInfAvg}}
\newcommand{\LimSupAvg}{\mathit{LimSupAvg}}
\newcommand{\LimSupAvgAutomat}[1]{\overline{#1}}
\newcommand{\LimInfAvgAutomat}[1]{\underline{#1}}

\newcommand{\InfAvgLan}[2]{\underline{#1}^{\geq #2}}
\newcommand{\SupAvgLan}[2]{\overline{#1}^{\geq #2}}

\newcommand{\VEC}[1]{\ensuremath{\overline{#1}}}

\newcommand{\Nat}{\ensuremath{\mathbb{N}}}

\begin{document}

\pagestyle{empty}

\def\One{\mathit{One}}
\makesavenoteenv{tabular}
\newcommand{\CMPSup}{\ensuremath{\bigwedge \textrm{MeanPayoffSup} }}
\newcommand{\CMPInf}{\ensuremath{\bigwedge \textrm{MeanPayoffInf} }}
\newcommand{\DMPSup}{\ensuremath{\bigvee \textrm{MeanPayoffSup} }}
\newcommand{\DMPInf}{\ensuremath{\bigvee \textrm{MeanPayoffInf} }}

\newcommand{\AppendixString}{the appendix}
\newcommand{\AppendixContent}[1]{#1}

\title{The Complexity of\\ Mean-Payoff Automaton Expression}
\author{Yaron Velner}
\institute{The Blavatnik School of Computer Science, Tel Aviv University, Israel}
\maketitle

\begin{abstract}
Quantitative languages are extension of Boolean languages that assign to each word a real number.
With quantitative languages, systems and specifications can be formalized more accurately. For example, a system may use a varying amount of some
resource (e.g., memory consumption, or power consumption) depending on its behavior, and a specification may assign a maximal amount of available resource to each behavior, or fix the long-run average available use of the resource.

Mean-payoff automata are finite automata with numerical weights on transitions that assign to each infinite path the long-run average of the transition weights.
Mean-payoff automata forms a class of quantitative languages that is not robust, since it is not closed under the basic algebraic operations: $\MIN$, $\MAX$, $\SUM$ and numerical complement.
The class of \emph{mean-payoff automaton expressions}, recently introduced by Chatterjee et al., is currently the only known class of quantitative languages that is robust, expressive and decidable.
This class is defined as the closure of mean-payoff automata under the basic algebraic operations.
In this work, we prove that all the classical decision problems for mean-payoff expressions are PSPACE-complete.
Our proof improves the previously known 4EXPTIME upper bound.
In addition, our proof is significantly simpler, and fully accessible to the automata-theoretic community.
\end{abstract}
\pagestyle{headings}
\section{Introduction}\label{sect:intro}
In algorithmic verification of reactive systems, the system is modeled as a finite-state transition system, and requirements are captured as languages of infinite words over system observations \cite{TemporalLogic1,TemporalLogic2}.
The classical verification framework only captures \emph{qualitative} aspects of system behavior, and in order to describe \emph{quantitative} aspects, for example, consumption of resources such as CPU and energy, the framework of \emph{quantitative languages} was proposed \cite{Weighted-Automaton}.

Quantitative languages are a natural generalization of Boolean languages that assign to every word a real number instead of a Boolean value.
With such languages, quantitative specifications can be formalized.
In this model, an implementation $L_A$ satisfies (or refines) a specification $L_B$ if $L_A(w) \leq L_B(w)$ for all words $w$.

This notion of refinement is a \emph{quantitative generalization of language inclusion}, and it can be used to check for example if for each behavior, the long-run average response time of the system lies below the specified average response requirement.
The other classical decision problems such as emptiness, universality, and language equivalence have also a natural quantitative extension.
For example, the \emph{quantitative emptiness problem} asks, given a quantitative language $L$ and a rational threshold $\nu$, whether there exists some word $w$ such that $L(w)\geq \nu$,
and the \emph{quantitative universality problem} asks whether $L(w)\geq \nu$ for all words $w$.
We also consider the notion of \emph{distance} between two quantitative languages $L_A$ and $L_B$, defined as $\sup_{w\in\Sigma^\omega} |L_A(w) - L_B(w)|$.

The model of mean-payoff automaton is a popular approach to express quantitative properties;
in this model, a \emph{payoff} (or a weight) is associated with each transition of the automaton, the mean-payoff of a finite run is simply the average of the payoff of the transitions in the run, and the mean-payoff of an infinite
run is the limit, as the length of the run tends to infinity.

In this work, we study the computational complexity of the classical decision problems for the class of quantitative languages that are defined by \emph{mean-payoff expression}.
An expression is either a deterministic\footnote{We note that the restriction to deterministic automata is inherent; for nondeterministic automata all the decision problems are undecidable~\cite{mean-payoff-Automaton-Expressions}.} mean-payoff automaton, or it is the $\MAX$, $\MIN$ or $\SUM$ of two mean-payoff expressions.
This class, introduced in \cite{mean-payoff-Automaton-Expressions}, is robust as it is closed under the $\MAX$, $\MIN$, $\SUM$ and the numerical complement operators \cite{mean-payoff-Automaton-Expressions}.

The decidability of the classical decision problems, as well as the computability of the distance problem, was first established in \cite{mean-payoff-Automaton-Expressions};
in this paper we describe alternative proofs for these results.
Our proofs offer the following advantages:
First, the proofs yield PSPACE complexity upper bounds, which match corresponding PSPACE lower bounds; in comparison to 4EXPTIME upper bounds achieved in \cite{mean-payoff-Automaton-Expressions}.
Second, our proofs reside only in the frameworks of graph theory and basic linear-programing,
which are common practices among the automata-theoretic community,
whereas a substantial part of the proofs in \cite{mean-payoff-Automaton-Expressions} resides in the framework of computational geometry.

Our proofs are based on a reduction from the emptiness problem to the feasibility problem for a set of linear inequalities;
for this purpose, inspired by the proofs in~\cite{NoisyInput}, we establish a connection between the emptiness problem and the problem of finding a multi-set of cycles, with certain properties, in a directed graph.
The reduction also reveals how to compute the maximum value of an expression, and therefore the decidability of all mentioned problems is followed almost immediately.

This paper is organized as follows:
In the next section we formally define the class of mean-payoff expressions;
in Section \ref{sect:2EXPTIMEAlgo} we describe a PSPACE algorithm that computes the maximum value of an expression that does not contain the $\MAX$ operator;
in Section \ref{sect:PSPACEforAllProblems} we show PSPACE algorithm for all the classical problems and prove their corresponding PSPACE lower bounds.
Due to lack of space, in some cases the proofs are omitted, and in other cases 
only sketches of the proofs are presented.
The full proofs are given in the appendix.

\section{Mean-Payoff Automaton Expression}\label{sect:MPExpression}
In this section we present the definitions of mean-payoff expressions from \cite{mean-payoff-Automaton-Expressions}.
\Heading{Quantitative languages.} A \emph{quantitative language} $L$ over a finite alphabet $\Sigma$ is a function $L:\Sigma^\omega \to \R$.
Given two quantitative languages $L_1$ and $L_2$ over $\Sigma$, we denote by $\MAX(L_1,L_2)$ (resp., $\MIN(L_1,L_2),$ $\SUM(L_1,L_2)$ and $-L_1$) the quantitative language that assigns $\MAX(L_1(w),L_2(w))$ (resp., $\MIN(L_1(w),L_2(w))$, $L_1(w) + L_2(w)$, and $-L_1(w)$) to each word $w\in\Sigma^\omega$.
The quantitative language $-L$ is called the \emph{complement} of $L$.

\Heading{Cut-point languages.} Let $L$ be a quantitative language over $\Sigma$.
Given a threshold $\nu\in \R$, the \emph{cut-point} language defined by $(L,\nu)$ is the language $L^{\geq \nu} = \Set{w\in\Sigma^\omega | L(w) \geq \nu}$. 

\Heading{Weighted automata.} A \emph{(deterministic) weighted automaton} is a tuple $A=\Tuple{Q,q_I,\Sigma,\delta,\WtFunc}$, where
(i)~$Q$ is a finite set of states, $q_I\in Q$ is the initial state, and $\Sigma$ is a finite alphabet; (ii)~$\delta \subseteq  Q\times \Sigma \times Q$ is a set of transitions such that for every $q\in Q$ and $\sigma\in\Sigma$ the size of the set $\Set{q'\in Q|(q,\sigma,q')\in\delta}$ is exactly 1; and (iii)~$\WtFunc : \delta \to \Q$ is a \emph{weight function}, where $\Q$ is the set of rationals.
\Comment{
We say that $A$ is \emph{deterministic} if for all $q\in Q$ and $\sigma\in\Sigma$, there exists $(q,\sigma,q')\in\delta$ for exactly one $q'\in Q$.
In this paper, we consider only deterministic automata.
In the sequel we show (by standard arguments) the complexity results that we present in this paper hold also for the case where the weights of the automata are rationals (rather than integers).}

\Heading{The product of weighted automata.}
The \emph{product} of the weighted automata $A_1,\dots,A_n$ such that $A_i = \Tuple{Q_i,q_I^i,\Sigma,\delta_i,\WtFunc_i}$ is the multidimensional weighted automaton $\Automat{A} = A_1 \times \dots \times A_n = \Tuple{Q_1 \times \dots \times Q_n, (q_I^1,\dots,q_I^n),\Sigma,\delta,\WtFunc}$ such that
$t=((q_1,\dots,q_n),\sigma,(q' _1,\dots,q' _n))\in\delta$ if $t_i = (q_i,\sigma,q' _i) \in \delta _i$ for all $i\in\RangeSet{1}{n}$,
and $\WtFunc(t) = (\Range{\WtFunc_1(t_1)}{\WtFunc_n(t_n)})\in\Q^n$.
We denote by $\Automat{A}_i$ the projection of the automaton $\Automat{A}$ to dimension $i$. 
\Heading{Words and runs.} A \emph{word} $w\in\Sigma^\omega$ is an infinite sequence of letters from $\Sigma$.
A \emph{run} of a weighted automaton $A$ over an infinite word $w=\sigma_1 \sigma_2 \dots$ is the (unique) infinite sequence $r = q_0 \sigma_1 q_1 \sigma_2 \dots$ of states and letters such that $q_0 = q_I$, and $(q_i,\sigma_{i+1},q_{i+1})\in\delta$ for all $i\geq 0$.
We denote by $\WtFunc(w)=\WtFunc(r)=v_0 v_1 \dots$ the sequence of weights that occur in $r$ where $v_i = \WtFunc(q_i,\sigma_{i+1},q_{i+1})$ for all $i\geq 0$.

\Heading{Quantitative language of mean-payoff automata.} The \emph{mean-payoff value} (or limit average) of a sequence $\overline{v}=v_0 v_1 \dots$ of real numbers is either\linebreak
$\LimInfAvg(\overline{v}) = \liminf_{n\to\infty} \frac{1}{n}\cdot \sum_{i=0}^{n-1} v_i$; or
$\LimSupAvg(\overline{v}) = \limsup_{n\to\infty} \frac{1}{n}\cdot \sum_{i=0}^{n-1} v_i$.
The quantitative language $\LimInfAvgAutomat{A}$ of a weighted automaton $A$ is defined by 
$\LimInfAvgAutomat{A}(w) = \LimInfAvg(\WtFunc(w))$;
analogously the quantitative language $\LimSupAvgAutomat{A}$ is defined by 
$\LimSupAvgAutomat{A}(w) = \LimSupAvg(\WtFunc(w))$.
In the sequel we also refer to the quantitative language $\LimInfAvgAutomat{A}$
as the \emph{$\LimInfAvg$ automaton $A$}, 
and analogously the \emph{$\LimSupAvg$ automaton $A$} is the quantitative language $\LimSupAvgAutomat{A}$.

\Heading{Mean-payoff automaton expressions.} A \emph{mean-payoff automaton expression} $E$ is obtained by the following grammar rule:
\begin{quote}
$E::= \LimInfAvgAutomat{A} | \LimSupAvgAutomat{A} | \MAX(E,E) | \MIN(E,E) | \SUM(E,E)$
\end{quote}
where $A$ is a \emph{deterministic} (one-dimensional) weighted automaton.
The quantitative language $L_E$ of a mean-payoff automaton expression $E$ is $L_E = \LimInfAvgAutomat{A}$ (resp., $L_E = \LimSupAvgAutomat{A}$) if
$E = \LimInfAvgAutomat{A}$ (resp., if $E = \LimSupAvgAutomat{A}$), and
$L_E = \OP(L_{E_1},L_{E_2})$ if $E = \OP(E_1,E_2)$ for $\OP \in \Set{\MAX,\MIN,\SUM}$.
We shall, by convenient abuse of notation, interchangeably use $E$ to denote both the expression and the quantitative language of the expression (that is, $E$ will also denote $L_E$).
An expression $E$ is called an \emph{atomic expression} if $E = \LimInfAvgAutomat{A}$ or $E= \LimSupAvgAutomat{A}$, where $A$ is a weighted automaton.

It was established in~\cite{mean-payoff-Automaton-Expressions} (and it follows almost immediately by the construction of the class) that the
class of mean-payoff automaton expressions is closed under $\MAX$, $\MIN$, $\SUM$ and numerical complement.

\Heading{Decision problems and distance.} We consider the following classical decision problems for a quantitative language defined by a mean-payoff expression.
Given a quantitative language $L$ and a threshold $\nu \in \Q$, the \emph{quantitative emptiness} problem asks whether there exists a word $w\in\Sigma^\omega$ such that $L(w)\geq \nu$, and the \emph{quantitative universality problem} asks whether $L(w)\geq \nu$ for all words $w\in\Sigma^\omega$.

Given two quantitative languages $L_1$ and $L_2$, the \emph{quantitative language-inclusion problem} asks whether $L_1(w)\leq L_2(w)$ for all words $w\in\Sigma^\omega$, and the \emph{quantitative language-equivalence problem} asks whether $L_1(w) = L_2(w)$ for all words $w\in\Sigma^\omega$.
Finally, the \emph{distance} between $L_1$ and $L_2$ is $D_{\sup} (L_1,L_2) = \sup_{w\in\Sigma^\omega} |L_1(w)-L_2(w)|$ ; and the corresponding computation problem is to compute the value of the distance.

\Heading{Maximum value of expression.}
Given an expression $E$, its \emph{supremum value} is the real number $\sup_{w\in\Sigma^\omega} E(w)$.
While it is obvious that such supremum exists, it was proved in~\cite{mean-payoff-Automaton-Expressions} that a maximum value also exists (that is,
there exists $w '\in\Sigma^\omega$ s.t $E(w ')=\sup_{w\in\Sigma^\omega} E(w)$).
Hence the \emph{maximum value} of the expression $E$ is $\sup_{w\in\Sigma^\omega} E(w)$ or equivalently $\max_{w\in\Sigma^\omega} E(w)$.

\Heading{Encoding of expressions and numbers.}
An expression $E$ is encoded by the tuple $( \langle E \rangle, \langle A_1 \rangle, \dots, \langle A_k \rangle)$,
where $\langle E \rangle$ is the expression string and $A_1,\dots,A_k$ are the weighted automata that occur in the expression,
w.l.o.g we assume that each automaton occur only once.
A rational number is encoded as a pair of integers, where every integer is encoded in binary.
\Comment{
We note that the complexity results that we present in this paper hold also for the case where the weights of the automata are rationals (rather than integers).
Indeed, if the automata have rational weights, then we multiply all the weights by a common divisor $N$, which is at most single-exponential in the weights (and therefore requires only polynomial number of bits to encode);
in the new expression $E'$ (which is formed by the new automata), for every infinite word $w$ we get that $E(w) = N\cdot E'(w)$, and thus, we can easily reduce all the decision problems to the case where the weights are integers.}

\section{PSPACE Algorithm for Computing the Maximum Value of $\MAX$-free Expressions}\label{sect:2EXPTIMEAlgo}
In this section we consider only $\MAX$-free expressions, which are expressions that contain only the $\MIN$ and $\SUM$ operators.
We will present a PSPACE algorithm that computes the maximum value of such expressions;
computing the maximum value amounts to computing the maximum threshold for which the expression is nonempty;
for this purpose we present four intermediate problems (and solutions), each problem is presented in a corresponding subsection below.
The first problem asks whether an intersection of cut-point languages of $\LimSupAvg$ automata is empty;
the second problem asks the same question for $\LimInfAvg$ automata;
the third problem asks if an arbitrary intersection of cut-point languages of $\LimSupAvg$ and $\LimInfAvg$ automata is empty;
and the last problem asks whether a $\MAX$-free expression is empty.
We will first present a naive solution for these problems;
the solution basically lists all the simple cycles in the product automaton of the automata that occur in the expression;
it then constructs linear constraints, with coefficients that depend on the weight vectors of the simple cycles, which their feasibility corresponds to the non-emptiness of the expression.

In the fifth subsection we will analyze the solution for the $\MAX$-free emptiness problem; we will show a PSPACE algorithm that solves the problem; and we will bound the number of bits that are needed to encode the maximum threshold for which the expression is nonempty (recall that such maximal threshold is the maximum value of the expression);
this will yield a PSPACE algorithm for computing the maximum value of a $\MAX$-free expression.

In subsections~\ref{subsect:First}-\ref{subsec:maxFreeNaive} we shall assume w.l.o.g that the product automaton of all the automata that occur in the expression is a strongly connected graph;
this can be done since in these subsections we do not refer to the complexity of the presented procedures.
\subsection{The emptiness problem for intersection of $\LimSupAvg$ automata}\label{subsect:First}
In this subsection we consider the problem where $k$ weighted automata $A_1,\dots,A_k$ and a rational threshold vector $\VEC{r} = (r_1,\dots,r_k)$ are given, and we need to decide whether there exists an infinite word $w\in\Sigma^\omega$ such that $\LimSupAvgAutomat{A_i}(w) \geq r_i$ for all $i\in\RangeSet{1}{k}$; equivalently, whether the intersection $\bigcap_{i=1}^k \SupAvgLan{A_i}{r_i}$ is nonempty.

Informally, we prove that there is such $w$ iff for every $i\in\RangeSet{1}{k}$, there is a word $w_i$ such that $\LimSupAvgAutomat{A_i}(w_i) \geq r_i$.

Formally, let $\Automat{A} = A_1 \times \dots \times A_k$ be the product automaton of the automata $A_1,\dots, A_k$.
Recall that an infinite word corresponds to an infinite path in $\Automat{A}$, and that w.l.o.g we assume that the graph of $\Automat{A}$ is strongly connected.
Let $C_1, C_2, \dots, C_n$ be the simple cycles that occur in $\Automat{A}$.
The next lemma claims that it is enough to find one cycle with average weight $r_i$ for every dimension $i$.
\begin{lem}\label{prop:LimSupDivideToDims}
There exists an infinite path $\pi$ in $\Automat{A}$ such that $\LimSupAvgAutomat{A_i}(\pi)\geq r_i$, for all $i\in\RangeSet{1}{k}$, iff for every $i\in\RangeSet{1}{k}$ there exists a simple cycle $C_i$ in $\Automat{A}$, with average weight at least $r_i$
\end{lem}
\begin{proof}
The direction from left to right is easy: 
Since for every $i\in\RangeSet{1}{k}$ there exists a path $\pi_i$ (namely $\pi$) such that $\LimSupAvgAutomat{A_i}(\pi_i) \geq r_i$ it follows that there exists a simple cycle in $\Automat{A}$ with average at least $r_i$ in dimension $i$.
(This fact is well-known for one-dimensional weighted automata, e.g., see \cite{Zwick}, and hence it is true for the projection of $\Automat{A}$ to the $i$-th dimension.)

For the converse direction, we assume that for every $i\in\RangeSet{1}{k}$ there exists a simple cycle $C_i$ in $\Automat{A}$ with average weight at least $r_i$.
Informally, we form the path $\pi$ by following the edges of the cycle $C_i$ until the average weight in dimension $i$ is sufficiently close to $r_i$, and then we do likewise for dimension $1 + (i\pmod k)$, and so on. (Full proof is given in \AppendixString.)
\pfbox
\end{proof}
Lemma \ref{prop:LimSupDivideToDims} shows that the emptiness problem for intersection of $\LimSupAvg$ automata can be naively solved by an exponential time algorithm that constructs the product automaton and checks if the desired cycles exist.

\subsection{The emptiness problem for intersection of $\LimInfAvg$ automata}\label{subsect:LimInf}
In this subsection we consider the problem where $k$ weighted automata $A_1,\dots,A_k$ and a rational threshold vector $\VEC{r} = (r_1,\dots,r_k)$ are given, and we need to decide whether there exists an infinite word $w\in\Sigma^\omega$ such that $\LimInfAvgAutomat{A_i}(w) \geq r_i$ for all $i\in\RangeSet{1}{k}$ ; or equivalently, whether the intersection $\bigcap_{i=1}^k \InfAvgLan{A_i}{r_i}$ is nonempty.

For $\LimInfAvg$ automata, the componentwise technique we presented in the previous subsection will not work; to solve the emptiness problem for the intersection of such automata we need the notion of $\VEC{r}$ multi-cycles.
\Heading{$\VEC{r}$ multi-cycles.}
Let $G$ be a directed graph equipped with a multidimensional weight function $\WtFunc:E\to \Q^k$, and let $\VEC{r}$ be a vector of rationals.
A \emph{multi-cycle} is a multi-set of simple cycles; the length of a multi-cycle $\mathcal{C} = \Set{C_1,\dots,C_n}$, denoted by $|\mathcal{C}|$, is $\sum_{i=1}^n |C_i|$.
A multi-cycle $\mathcal{C} = \Set{C_1,\dots,C_n}$ is said to be an \emph{$\VEC{r}$ multi-cycle} if $\frac{1}{|\mathcal{C}|} \sum_{j=1}^n \WtFunc(C_j) \geq \VEC{r}$, that is, if the average weight of the multi-cycle, in every dimension $i$, is at least $r_i$.

In the sequel we will establish a connection between the problem of finding an $\VEC{r}$ multi-cycle and the emptiness problem for intersection of $\LimInfAvg$ automata.

A polynomial time algorithm that decides if an $\VEC{r}$ multi-cycle exists is\linebreak known~\cite{ZeroCircut};
in this work however, it is sufficient to present the naive way for finding such multi-cycles;
for this purpose we construct the following set of linear constraints:
Let $\mathbb{C}$ denote the set of all simple cycles in $\Automat{A}$;
for every $c\in \mathbb{C}$ we define a variable $X_c$; we define the \emph{$\VEC{r}$ multi-cycle constraints} to be:
\begin{quote}
$\sum_{c\in \mathbb{C}} X_c \WtFunc(c) \geq \VEC{r}$ ; 
$\sum_{c\in \mathbb{C}} |c|X_c = 1$ ; and for every $c\in\mathbb{C}$: $X_c \geq 0$\end{quote}
In the next lemma we establish the connection between the feasibility of the $\VEC{r}$ multi-cycle constraints and the existence of an $\VEC{r}$ multi-cycle.
\begin{lem}\label{prop:MultiCycleIffConstrains}
The automaton $\Automat{A}$ has an $\VEC{r}$ multi-cycle iff the corresponding $\VEC{r}$ multi-cycle constraints are feasible.
\end{lem}
\begin{proof}
The direction from left to right is immediate,
indeed if we define $X_c$ as the number of occurrences of cycle $c$ in the witness $\VEC{r}$ multi-cycle divided by the length of that multi-cycle, then we get a solution for the set of constraints.

In order to prove the converse direction, it is enough to notice that if the constraints are feasible then they have a rational solution.
Let $\VEC{X}$ be such rational solution, and let $N$ be the least common multiple of all the denominators of the elements of $\VEC{X}$;
by definition, the multi-set that contains $N X_c$ copies of the cycle $c$ is an $\VEC{r}$ multi-cycle. \pfbox
\end{proof}

In the following lemma we establish a connection between the problem of finding an $\VEC{r}$ multi-cycle and the emptiness problem for intersection of $\LimInfAvg$ automata.
\begin{lem}\label{lem:ConnectionBetweenMultiCycleAndEmptiness}
There exists an infinite path $\pi$ in $\Automat{A}$ such that $\LimInfAvgAutomat{A_i}(\pi)\geq r_i$, for all $i\in\RangeSet{1}{k}$, iff the graph of $\Automat{A}$ contains 
an $\VEC{r} = (r_1,\dots,r_k)$ multi-cycle.
\end{lem}
\begin{proof}
To prove the direction from right to left, we show, in the following lemma, that if an $\VEC{r}$ multi-cycle does not exist, then for every infinite path there is a dimension $i$ for which $\LimInfAvgAutomat{A_i}(\pi) < r_i$.
\begin{lem}\label{prop:NoMultiCycleImpliesNegPayoff}
Let $G=(V,E)$ be a directed graph equipped with a weight function $\WtFunc:E\to\Q^k$,
and let $\VEC{r} \in \Q^k$ be a threshold vector.
If $G$ does not have an $\VEC{r}$ multi-cycle, then there exist constants $\epsilon_G > 0$ and $m_G \in \Nat$ such that for every finite path $\pi$ there is a dimension $i$ for which $\WtFunc_i(\pi) \leq m_G + (r_i - \epsilon_G) |\pi|$.
\end{lem}
\Appendix{
\begin{proof}
Let $\epsilon_G$ be the minimal $\epsilon$ for which the $\VEC{r}-\epsilon$ multi-cycle constraints are feasible;
note that $\epsilon_G$ is the optimal solution for a linear programming problem;
the constraints of the linear programming problem are feasible, since the $\VEC{r}-\epsilon$ multi-cycle constraints are feasible for $\epsilon = +2W$;
in addition the solution is bounded by $\epsilon = -W$;
hence such $\epsilon_G$ must exist, and if $G$ does not have an $\VEC{r}$ multi-cycle then $\epsilon_G > 0$ (due to Lemma~\ref{prop:MultiCycleIffConstrains}).

Let $\pi$ be an arbitrary finite path in $G$, we decompose $\pi$ into three paths namely $\pi_0, \pi_c$ and $\pi_1$ such that $|\pi_0|, |\pi_1| \leq |V|$ and $\pi_c$ is a cyclic path (this can be done since any path longer then $|V|$ contains a cycle).
Let $C_1, \dots, C_n$ be the simple cycles that occur in $\pi_c$, and let $m_i$ be the number of occurrences of cycle $C_i$ in $\pi_c$.
By definition we get that $\frac{1}{|\pi_c|} \WtFunc(\pi_c) = \frac{1}{|\pi_c|} \sum_{j=1}^n m_j \WtFunc(C_j)$.
Towards contradiction let us assume that there exists $\delta < \epsilon_G$ such that for every dimension $\frac{1}{|\pi_c|} w_i(\pi_c) \geq r_i - \delta$.
Hence, by definition the $\VEC{r}-\delta$ multi-cycle constraints are feasible, which contradicts the minimality of $\epsilon_G$.

Therefore, for $m_G = -2|V|W$ we get that for every finite path $\pi$ there exists a dimension $i$ for which $\WtFunc_i(\pi) \leq m_G + (r_i - \epsilon_G)|\pi|$. \pfbox
\end{proof}
}
Lemma~\ref{prop:NoMultiCycleImpliesNegPayoff} implies that if an $\VEC{r}$ multi-cycle does not exists, then for every infinite path $\pi$ there exist a dimension $i$ and an infinite sequence of indices $j_1 < j_2 < j_3 \dots$ such that the average weight of the prefix of $\pi$, of length $j_m$, is at most $r_i -\frac{\epsilon_G}{2}$, for all $m\in\Nat$.
Hence by definition $\LimInfAvg_i(\pi) < r_i$.

In order to prove the converse direction, let us assume that $G$ has an $\VEC{r}$ multi-cycle $\mathcal{C} = {C_1, \dots, C_n}$, such that the cycle $C_i$ occurs $m_i$ times in $\mathcal{C}$.
We obtain the witness path $\pi$ in the following way (we demonstrate the claim for $n = 2$):
let $\pi_{12}$ be a path from $C_1$ to $C_2$ and $\pi_{21}$ be a path from $C_2$ to $C_1$ (recall that the graph is strongly connected), we define 
\begin{quote}
$\pi = C_1 ^{m_1} \pi_{12} C_2 ^{m_2} \pi_{21} (C_1 ^{m_1})^2 \pi_{12} (C_2 ^{m_2})^2 \pi_{21} \dots (C_1 ^{m_1})^\ell \pi_{12} (C_2 ^{ m_2})^\ell \pi_{21} \dots$
\end{quote}
Informally, the long-run average weight of the path $\pi$ is determined only by the cycles $C_1^{m_1}$ and $C_2^{m_2}$, since the effect of the paths $\pi_{12}$ and $\pi_{21}$ on the average weight of a prefix of $\pi$ becomes negligible as the length of the prefix tends to infinity.
Thus $\LimInfAvgAutomat{A_i}(\pi)\geq r_i$ for every dimension $i$, which concludes the proof of Lemma~\ref{lem:ConnectionBetweenMultiCycleAndEmptiness} \pfbox
\end{proof}

Lemma~\ref{lem:ConnectionBetweenMultiCycleAndEmptiness} and Lemma~\ref{prop:MultiCycleIffConstrains} immediately give us the following naive algorithm for the emptiness problem for intersection of $\LimInfAvg$ automata:
First, construct the product automaton; second, list all the simple cycles in the product automaton; third, construct the $\VEC{r}$ multi-cycle constraints and check for their feasibility.

When the automata and the threshold vector are clear from the context,
we shall refer to the $\VEC{r}$ multi-cycle constraints, which are constructed from the intersection of the given $\LimInfAvg$ automata and the threshold $\VEC{r}$, as the \emph{lim-inf constraints}.

\subsection{The emptiness problem for intersection of $\LimSupAvg$ and $\LimInfAvg$ automata}\label{subsect:LimInfAndSup}
In this subsection we consider the problem where $2k$ weighted automata\linebreak $A_1,\dots,A_k$, $B_1, \dots, B_k$ and two $k$-dimensional rational threshold vectors $\VEC{r^a}$ and $\VEC{r^b}$ are given, and we need to decide whether there exists an infinite word $w\in\Sigma^\omega$ such that $\LimInfAvgAutomat{A_i}(w) \geq r^a_i$ and
$\LimSupAvgAutomat{B_i}(w) \geq r^b_i$ for all $i\in\RangeSet{1}{k}$;
or equivalently, whether the intersection
$(\bigcap_{i=1}^k \InfAvgLan{A_i}{r^a_i}) \cap 
(\bigcap_{i=1}^k \SupAvgLan{B_i}{r^b_i})$ is nonempty.

Our solution will be a result of the following two lemmata.
The first lemma claims that there is a word that satisfies all the conditions iff there are words $w_1,\dots,w_k$ such that $w_j$ satisfies all the lim-inf conditions and the lim-sup condition for the automaton $B_j$.
\begin{lem}\label{prop:SupInfAvgDivideAndConcore}
There exists an infinite word $w$ for which $\LimInfAvgAutomat{A_i}(w) \geq r^a_i$ and
$\LimSupAvgAutomat{B_i}(w) \geq r^b_i$ for all $i\in\RangeSet{1}{k}$
iff
there exist $k$ infinite words $w_1, w_2, \dots w_k$ such that for every $j\in\RangeSet{1}{k}$:
\begin{quote}
$\LimInfAvgAutomat{A_i}(w_j) \geq r^a_i$ for all $i\in\RangeSet{1}{k}$ ;
and
$\LimSupAvgAutomat{B_j}(w_j) \geq r^b_j$
\end{quote}
\end{lem}
The second lemma shows that the emptiness problem for an intersection of lim-inf automata and one lim-sup automaton can be reduced to the emptiness problem for an intersection of lim-inf automata.
\begin{lem}\label{prop:ConjSupAndInfBecomesInf}
The intersection $\SupAvgLan{B_1}{r^b_1} \cap (\bigcap_{i=1}^k \InfAvgLan{A_i}{r^a_i})$ is nonempty iff the intersection 
$\InfAvgLan{B_1}{r^b_1} \cap (\bigcap_{i=1}^k \InfAvgLan{A_i}{r^a_i})$ is nonempty.
\end{lem}
Due to Lemma~\ref{prop:SupInfAvgDivideAndConcore} and~\ref{prop:ConjSupAndInfBecomesInf} we can solve the emptiness problem for intersection of lim-inf and lim-sup automata in the following way:
First, we construct the product automata $\Automat{A}^i = A_1 \times \dots \times A_k \times B_i$ for all $i\in\RangeSet{1}{k}$
 and list all the simple cycles that occur in it;
second, we construct the threshold vector $\VEC{r^i} = (r^a_1,\dots,r^a_k,r^b_i)$ and check if the graph of $\Automat{A}^i$ has an $\VEC{r^i}$ multi-cycle, for all $i\in\RangeSet{1}{k}$.
Due to Lemma~\ref{prop:SupInfAvgDivideAndConcore} and~\ref{prop:ConjSupAndInfBecomesInf} the intersection is nonempty iff every $\Automat{A}^i$ has an $\VEC{r^i}$ multi-cycle, that is, if the $\VEC{r^i}$ multi-cycle constraints are feasible.

Recall that the existence of an $\VEC{r^i}$ multi-cycle in the graph of $\Automat{A}^i$ is equivalent to the feasibility of the corresponding lim-inf constraints for $\Automat{A}^i$ and $\VEC{r^i}$;
in the sequel, we will refer to the set of constraints $\bigcup_{i=1}^k\Set{\mbox{lim-inf constraints for $\Automat{A}^i$ and $\VEC{r^i}$}}$ as the \emph{$\MIN$-only constraints}.
(As we use them to decide the emptiness of expressions that contain only the $\MIN$ operator.)

In this subsection we proved that the emptiness of the intersection of\linebreak $\LimInfAvg$ and $\LimSupAvg$ automata is equivalent to the feasibility of the corresponding $\MIN$-only constraints.

\subsection{The emptiness problem for $\MAX$-free expressions}\label{subsec:maxFreeNaive}
In this subsection we solve the emptiness problem for $\MAX$-free expressions.
The solution we present is a reduction to the emptiness problem for an intersection of lim-inf and lim-sup automata with a threshold vector that satisfies certain linear constraints;
the reduction yields a naive double-exponential complexity upper-bound for the problem, 
which we will improve in the succeeding subsection.

The reduction is based on the next simple observation.
\begin{Obs}\label{obs:SimpleObs}
The expression $E = E_1 + E_2$ is nonempty with respect to the rational threshold $\nu$ iff there exist two thresholds $\nu_1,\nu_2\in\R$ such that (i)~The intersection of the cut-point languages $E_1^{\geq \nu_1}$ and $E_2^{\geq \nu_2}$ is nonempty; and (ii)~~$\nu_1 + \nu_2 \geq \nu$. 
\end{Obs}
If $E = E_1 + E_2$ and $E_1$ and $E_2$ are $\MIN$-only expressions then we 
decide the emptiness of $E$ in the following way:
we combine the $\MIN$-only constraints for the expressions $E_1$ and $E_2$ with respect to arbitrary thresholds $r_1$ and $r_2$ (that is, $r_1$ and $r_2$ are variables in the constraints), note that these are still linear constraints;
we then check the feasibility of the constraints subject to $r_1 + r_2 \geq \nu$.
(Note that as all the constraints are linear, this can be done by linear programming.)

The next lemma shows that in the general case, the emptiness problem for an arbitrary $\MAX$-free expression and a threshold $\nu$ can be reduced, in polynomial time, to the emptiness problem for an intersection of lim-inf and lim-sup automata with respect to threshold vectors $r^a$ and $r^b$ subject to certain linear constraints on $r^a$ and $r^b$.

\begin{lem}\label{prop:ReductionFromMaxFreeToConjunction}
Let $E$ be a $\MAX$-free expression with atomic expressions $e_1,\dots,e_k$, and let $\nu$ be a rational threshold, then there exist a $2k\times 2k$ matrix $M_{E}$ and a $2k$-dimensional vector $\VEC{b_\nu}$, with rational coefficients, and computable in polynomial time (from $E$ and $\nu$) such that:
\begin{quote}
The expression $E$ is nonempty (with respect to $\nu$) iff
there exists a $2k$-dimensional vector of reals $\VEC{r}$ such that the intersection $\bigcap_{i=1}^k e_i^{\geq r_i}$ is nonempty \textbf{and} $M_{E}\times \VEC{r} \geq \VEC{b_\nu}$.
\end{quote}
\end{lem}
Instead of formally proving the correctness of Lemma~\ref{prop:ReductionFromMaxFreeToConjunction}, we provide a generic example that illustrates the construction of the matrix $M_{E}$ and the vector $\VEC{b_\nu}$.
\begin{examp} \label{examp:RemoveSum}
Let $E = \MIN(\LimInfAvgAutomat{A_1},(\LimInfAvgAutomat{A_2}+\LimSupAvgAutomat{A_3}))+
\MIN(\LimSupAvgAutomat{A_4},\LimInfAvgAutomat{A_5})$.
Then for every $\nu\in\R$, each the following condition is equivalent to $E ^{\geq \nu} \neq \emptyset$.
\begin{itemize}
\item $\exists r_6,r_7 \in \R$ such that 
$L_{\MIN(\LimInfAvgAutomat{A_1},(\LimInfAvgAutomat{A_2}+\LimSupAvgAutomat{A_3}))}^{\geq r_6} \cap 
L_{\MIN(\LimSupAvgAutomat{A_4},\LimInfAvgAutomat{A_5})}^{\geq r_7} \neq \emptyset$
\textbf{and} $r_6 + r_7 \geq \nu$.
\item $\exists r_1, r_4, r_5, r_6, r_7, r_8$ such that
$\LimInfAvgAutomat{A_1}^{\geq r_1} \cap L_{\LimInfAvgAutomat{A_2}+\LimSupAvgAutomat{A_3}}^{\geq r_8} \cap \LimSupAvgAutomat{A_4}^{\geq r_4} \cap \LimInfAvgAutomat{A_5}^{\geq r_5}\neq \emptyset$ \textbf{and}
$r_1 \geq r_6$, $r_8 \geq r_6$, $r_4 \geq r_7$, $r_5 \geq r_7$ and $r_6 + r_7 \geq \nu$.
\item $\exists r_1, r_2, r_3, r_4, r_5, r_6, r_7, r_8$ such that
$\LimInfAvgAutomat{A_1}^{\geq r_1} \cap \LimInfAvgAutomat{A_2}^{\geq r_2}\cap \LimSupAvgAutomat{A_3}^{\geq r_3} \cap \LimSupAvgAutomat{A_4}^{\geq r_4} \cap \LimInfAvgAutomat{A_5}^{\geq r_5}\neq \emptyset$ \textbf{and}
$r_1 \geq r_6$, $r_2 + r_3 \geq r_8$,
$r_8 \geq r_6$, $r_4 \geq r_7$, $r_5 \geq r_7$ and $r_6 + r_7 \geq \nu$.
\end{itemize}
The reader should note that we associate every variable $r_i$ either with a sub-expression or with an atomic expression;
as we assume that each atomic expression occurs only once,
the number of variables is at most $2k$.
\end{examp}
Hence we can solve the emptiness problem for a rational threshold $\nu$ and a $\MAX$-free expression $E$, which contains the atomic expressions $e_1,\dots,e_k$ in the following way:
First, we construct the matrix $M_{E}$ and the vector $\VEC{b_\nu}$;
second, we construct the $\MIN$-only constraints for the intersection $\bigcap_{i=1}^k e_i ^{\geq r_i}$ and check for their feasibility subject to the constraints $M_{E} \times r \geq \VEC{b_\nu}$.

In the sequel we will refer to the $\MIN$-only constraints along with the $M_{E,\nu} \times \VEC{r} \geq \VEC{b_\nu}$ constraints as the \emph{$\MAX$-free constraints};
we will show that even though the size of the constraints is double-exponential, there is a PSPACE algorithm that decides their feasibility (when the input is $E$ and $\nu$).

\subsection{PSPACE algorithm for the emptiness problem of $\MAX$-free expressions}\label{subsect:maxFreePSPACE}
In this subsection we will present a PSPACE algorithm that for given $\MAX$-free expression $E$ and rational threshold $\nu$, decides the feasibility of the $\MAX$-free constraints;
as shown in subsection~\ref{subsec:maxFreeNaive}, such algorithm also solves the emptiness problem for $\MAX$-free expressions.
Informally, we will show that if the $\MAX$-free constraints are feasible then they have a \emph{short} solution, and that a short solution can be verified by a polynomial-space machine;
hence the problem is in NPSPACE, and due to Savitch Theorem, also in PSPACE.

The next lemma describes key properties of $\MAX$-free constraints, which we will use to obtain the PSPACE algorithm.
\begin{lem}\label{prop:PropsOfMaxFreeConstrains}
For every $\MAX$-free expression $E$:
\begin{enumerate}
\item For every threshold $\nu$, the $\MAX$-free constraints have at most $O(k^2)$ constraints, where $k$ is the number of automata that occur in $E$, that are not of the form of $x \geq 0$, where $x$ is a variable.
\item There exists a bound $t$, polynomial in the size of the expression, such that for every threshold $\nu$, the $\MAX$-free constraints are feasible iff there is a solution that assigns a nonzero value to at most $t$ variables.
\item There exists a bound $t$, polynomial in the size of the expression, such that the maximum threshold $\nu\in\R$, for which the $\MAX$-free constraints are feasible, is a rational and can be encoded by at most $t$ bits. (In particular such maximum $\nu$ exists.)
\end{enumerate}
\end{lem}
Recall that a rational solution for the $\MAX$-free constraints corresponds to vectors of thresholds and a set of multi-cycles, each multi-cycle with an average weight that matches its corresponding threshold vector;
by Lemma~\ref{prop:PropsOfMaxFreeConstrains}(2) the number of different simple cycles that occur in the witness multi-cycles set is at most $t$.
We also observe that if a multi-set of cycles (that are not necessarily simple) with average weight vector $\VEC{\nu}$ exists, then a $\VEC{\nu}$ multi-cycle (of simple cycles) also exists, since we can decompose every non-simple cycle to a set of simple cycles;
thus, a $\VEC{\nu}$ multi-cycle exists iff there exists a multi-set of \emph{short} cycles, where the length of each cycle in the multi-set is at most the number of vertices in the graph (note that in particular, every simple cycle is short).

Hence, we can decide the feasibility of the $\MAX$-free constraints in the following way:
First, we guess $t$ weight vectors of $t$ short cycles that occur in the same strongly connected component (SCC) of the product automaton of all the automata that occur in the expression;
second, we construct the $O(k^2)$ constraints of the $\MAX$-free constraints and assign zero values to all the variables of the non-chosen cycles;
third, we check the feasibility of the formed $O(k^2)$ constraints, where each constraint has at most $t + 1$ variables.
\Comment{
 we construct the product automaton of all automata occur in the expression;
second, we list all the short cycles that occur in the product automaton;
third, we choose, nondeterministically, $t$ cycles, all in the same SCC of the product automata, and list their average weight vectors;
fourth, we construct the $O(k^2)$ constraints of the $\MAX$-free constraints and assign zero values to all the variables of the non-chosen cycles;
fifth, we check the feasibility of the formed $O(k^2)$ constraints, where each constraint has at most $t + 1$ variables.}

Note that we can easily perform the last two steps in polynomial time (as the values of the weight vector of every short cycle can be encoded by polynomial number of bits);
hence, to prove the existence of a PSPACE algorithm, it is enough to show
how to encode (and verify by a polynomial-space machine) $t$ average weight vectors of $t$ short cycles that belong to one SCC of the product automaton.
Informally, the encoding scheme is based on the facts that every vertex in the product automaton is a $k$-tuple of states, and that a path is a sequence of alphabet symbols;
the verification is done by simulating the $k$ automata in parallel, and since the size of the witness string should be at most exponential, we can do it with a polynomial-size tape. (More details are given in \AppendixString.)

To conclude, we proved that there is a PSPACE algorithm that decides the feasibly of the $\MAX$-free constraints, and therefore the next lemma follows.
\begin{lem}\label{lem:MaxFreeEmptInPSPACE}
The emptiness problem for \Comment{the class of }$\MAX$-free expressions is in PSPACE.
\end{lem}

Lemma~\ref{lem:MaxFreeEmptInPSPACE} along with Lemma~\ref{prop:PropsOfMaxFreeConstrains}(3) imply a PSPACE algorithm that computes the maximum value of a $\MAX$-free expression;
the next lemma formally states this claim.
\begin{lem}\label{lem:MaxFreeMaxInPSPACE}
(i)~The maximum value of a $\MAX$-free expression is a rational value that can be encoded by polynomial number of bits (in particular, every expression has a maximum value); and
(ii)~The maximum value of a $\MAX$-free expression is PSPACE computable.
\end{lem}

\section{The Complexity of Mean-Payoff Expression Problems}\label{sect:PSPACEforAllProblems}
In this section we will prove PSPACE membership, and PSPACE hardness, for the classical mean-payoff expression problems;
the key step in the proof of the PSPACE membership is the next theorem, which extends Lemma~\ref{lem:MaxFreeMaxInPSPACE} to arbitrary expressions (as opposed to only $\MAX$-free expressions).
\begin{thm}\label{lem:MaxForExpression}
(i)~The maximum value of an expression is a rational value that can be encoded by polynomial number of bits (in particular, every expression has a maximum value); and
(ii)~The maximum value of an expression is PSPACE computable.
\end{thm}
\begin{proof}[of Theorem~\ref{lem:MaxForExpression}]
Informally, we prove that if the number of $\MAX$ operators in the expression $E$ is $m > 0$, then we can construct in linear time two expressions $E_1$ and $E_2$, each with at most $m-1$ $\MAX$ operators and of size at most $|E|$, such that $E = \MAX(E_1,E_2)$;
hence, in order to compute the maximum value of $E$, we recursively compute the maximum values of $E_1$ and $E_2$, and return the maximum of the two values;
note that if the expression is $\MAX$-free (that is, if $m=0$),
then thanks to Lemma~\ref{lem:MaxFreeMaxInPSPACE}, the maximum value is PSPACE computable and can be encoded by polynomial number of bits.
(Formal proof is given in \AppendixString.) \pfbox
\end{proof}
The PSPACE membership of the classical problems follows almost trivially from 
Theorem~\ref{lem:MaxForExpression}.
Indeed,
for a given threshold, an expression is empty if its maximum value is less than the threshold,
and an expression is universal if its minimum value (that is, the maximum of its numerical complement) is not less than the threshold;
the language inclusion and equivalence problems are special cases of the universality problem (since the class of mean-payoff expressions is closed under numerical complement and the $\SUM$ operator);
and the distance of the expressions $E_1$ and $E_2$ is the maximum value of the expression $F = \MAX(E_1 - E_2, E_2 - E_1)$.

The PSPACE lower bounds for the decision problems are obtained by reductions from the emptiness problem for intersection of regular languages (see proofs in \AppendixString), which is PSPACE-hard~\cite{IntersectionOfRegularLanguagesIsPSPACEComplete}.

Thus, we get the main result of this paper:
\begin{thm}\label{thm:MainThm}
For the class of mean-payoff automaton expressions, the quantitative
emptiness, universality, language inclusion, and equivalence problems are PSPACE-complete, and the distance is PSPACE computable.
\end{thm}

\section{Conclusion and Future Work}
We proved tight complexity bounds for all classical decision problems for mean-payoff expressions and for the distance computation problem.
Future work is to investigate the decidability of games with mean-payoff expression winning condition.
\Heading{Acknowledgements.}
The author would like to thank Prof. Alexander Rabinovich for his helpful comments.
This research was partially supported by the Israeli Centers of Research Excellence (I-CORE) program, (Center  No. 4/11).

\nocite{*}

\bibliographystyle{plain}	
\bibliography{mpbibtex}
\AppendixContent{
\newpage
\appendix
\section*{Appendix}
\section{Proofs for Section~\ref{sect:2EXPTIMEAlgo}}
\subsection{Proofs for subsection~\ref{subsect:First}}
\subsubsection{Proof of Lemma~\ref{prop:LimSupDivideToDims}}
\begin{proof}
The direction from left to right is easy:
 
Since for every $i\in\RangeSet{1}{k}$ there exists a path $\pi_i$ (namely $\pi$) such that $\LimSupAvgAutomat{A_i}(\pi_i) \geq r_i$ it follows that there exists a simple cycle in $\Automat{A}$ with average weight at least $r_i$ in dimension $i$.
(This fact is well-known for one-dimensional weighted automata (e.g., see \cite{Zwick}), and hence it is true for the projection of $\Automat{A}$ to the $i$-th dimension.)

For the converse direction, we assume that for every $i\in\RangeSet{1}{k}$ there exists a simple cycles $C_i$ in $\Automat{A}$ with average weight at least $r_i$.
Informally, we form the path $\pi$ by following the edges of the cycle $C_i$ until the average weight in dimension $i$ is sufficiently close to $r_i$, and then we do likewise for dimension $1 + (i\pmod k)$, and so on.
Formally, 
for every $\epsilon > 0$, and an arbitrary finite path $\lambda$ in $\Automat{A}$, we construct the path $\pi^\epsilon(\lambda)$ in the following way:
The first part of the path is $\lambda$, then we continue to a vertex in the cycle $C_1$ and follow the edges of the cycle $C_1$ until the average weight in the first dimension of the path is at least $r_1 - \epsilon$;
we then continue to a vertex in $C_2$ and follow the edges of $C_2$ until the average weight in the second dimension is at least $r_2 - \epsilon$;
we repeat the process also for $C_3, \dots , C_k$.
We recall that $\Automat{A}$ is strongly connected, and therefore $\pi^\epsilon(\lambda)$ is a valid path.
Note that for every $i\in\RangeSet{1}{k}$ there is a prefix of $\pi^\epsilon(\lambda)$ with average weight at least $r_i - \epsilon$ in dimension $i$.
Let $\pi^0$ be an arbitrary path, let $\epsilon_i = \frac{1}{2^i}$ and let $\pi^i = \pi^{\epsilon_i}(\pi^{i-1})$.
The reader can verify that the infinite path $\pi = \pi^0 \pi^1 \pi^2 \dots$ satisfies $\LimSupAvgAutomat{A_i}(\pi) \geq r_i$ for every $i\in\RangeSet{1}{k}$, which concludes the proof of the lemma.\pfbox
\end{proof}

\subsection{Proofs for subsection~\ref{subsect:LimInf}} 
\subsubsection{Proof of Lemma~\ref{lem:ConnectionBetweenMultiCycleAndEmptiness}}
\begin{proof}
To prove the direction from right to left, we show, in the following lemma, that if an $\VEC{r}$ multi-cycle does not exist, then for every infinite path there is a dimension $i$ such that $\LimInfAvgAutomat{A_i}(\pi) < r_i$.
\begin{lem}\label{appendx:prop:NoMultiCycleImpliesNegPayoff}
Let $G=(V,E)$ be a directed graph equipped with a weight function $\WtFunc:E\to\Q^k$,
and let $\VEC{r} \in \Q^k$ be a threshold vector.
If $G$ does not have an $\VEC{r}$ multi-cycle, then there exist constants $\epsilon_G > 0$ and $m_G \in \Nat$ such that for every finite path $\pi$ there is a dimension $i$ for which $\WtFunc_i(\pi) \leq m_G + (r_i - \epsilon_G) |\pi|$.
\end{lem}
\begin{proof}
Let $\epsilon_G$ be the minimal $\epsilon$ for which the $\VEC{r}-\epsilon$ multi-cycle constraints are feasible;
note that $\epsilon_G$ is the optimal solution for a linear programming problem;
the constraints of the linear programming problem are feasible, since the $\VEC{r}-\epsilon$ multi-cycle constraints are feasible for $\epsilon = 2W + \max\Set{r_1,\dots,r_k}$ (where $W$ is the maximal weight that occur in the graph);
in addition the solution is bounded by $\epsilon = -W$;
hence such $\epsilon_G$ must exist, and if $G$ does not have an $\VEC{r}$ multi-cycle then $\epsilon_G > 0$ (due to Lemma~\ref{prop:MultiCycleIffConstrains}).

Let $\pi$ be an arbitrary finite path in $G$ of length longer than $|V|$, we decompose $\pi$ into three paths namely $\pi_0, \pi_c$ and $\pi_1$ such that $|\pi_0|, |\pi_1| \leq |V|$ and $\pi_c$ is a cyclic path (this can be done since any path longer then $|V|$ contains a cycle).
Let $C_1, \dots, C_n$ be the simple cycles that occur in $\pi_c$, and let $m_i$ be the number of occurrences of cycle $C_i$ in $\pi_c$.
By definition we get that $\frac{1}{|\pi_c|} \WtFunc(\pi_c) = \frac{1}{|\pi_c|} \sum_{j=1}^n m_j \WtFunc(C_j)$; and $|\pi_c| = \sum_{j=1}^n m_j|C_j|$.
Towards contradiction let us assume that there exists $\delta < \epsilon_G$ such that for every dimension $\frac{1}{|\pi_c|} w_i(\pi_c) \geq r_i - \delta$.
Hence, by definition the $\VEC{r}-\delta$ multi-cycle constraints are feasible, which contradicts the minimality of $\epsilon_G$.

Therefore, for $m_G = -2|V|W$ (where $-W$ is the minimal weight that occur in the graph) we get that for every finite path $\pi$ there exists a dimension $i$ for which $\WtFunc_i(\pi) \leq m_G + (r_i - \epsilon_G)|\pi|$. \pfbox
\end{proof}

Lemma~\ref{prop:NoMultiCycleImpliesNegPayoff} implies that if an $\VEC{r}$ multi-cycle does not exists, then for every infinite path $\pi$ there exist a dimension $i$ and an infinite sequence of indices $j_1 < j_2 < j_3 \dots$ such that the average weight of the prefix of $\pi$ of length $j_m$ is at most $r_i -\frac{\epsilon_G}{2}$, for all $m\in\Nat$.
Hence by definition $\LimInfAvg_i(\pi) < r_i$.

In order to prove the converse direction, let us assume that $G$ has an $\VEC{r}$ multi-cycle $\mathcal{C} = {C_1, \dots, C_n}$, such that the cycle $C_i$ occurs $m_i$ times in $\mathcal{C}$.
We obtain the witness path $\pi$, for which we will prove that $\LimInfAvg(\pi)\geq \VEC{r}$, in the following way:
Let $\pi_{i \to j}$ be a path from the cycle $C_i$ to the cycle $C_j$ (recall that the graph is strongly connected).
For every index $\ell\in\Nat$ we define the finite path $\pi^\ell$ to be
\begin{quote}
$(C_1 ^ {m_1}) ^ \ell \pi_{1\to 2} (C_2 ^ {m_2}) ^ \ell \pi_{2\to 3} \dots
\pi_{n-1\to n} (C_n ^{m_n})^\ell \pi_{n\to 1}$
\end{quote}
We set $\pi = \pi^1 \pi^2 \dots \pi^i \dots$, and claim that $\LimInfAvg_i(\pi)$ is at least $r_i$ in every dimension $i$;
for this purpose, it is enough to prove that for every $\epsilon > 0$ we have $\LimInfAvg_i(\pi) \geq r_i -\epsilon$.
For the rest of the proof we shall assume w.l.o.g that $r_i = 0$.
Let us denote by $\pi_j$ the finite path $\pi^1 \pi^2 \dots \pi^j$;
by $P$ the sum $|\pi_{1\to 2}| + |\dots+ |\pi_{i \to i+1}| + |\dots + \pi_{n-1 \to n}| + |\pi_{n \to 1}|$;
by $|\mathcal{C}|$ the total length of the multi-cycle, that is, $\sum_{i=1}^n m_i|C_i|$;
and by $-W$ the minimal weight that occur in the graph (in all dimensions).
Then the average weight of dimension $i$ for the path $\pi_j$ is at least
\begin{equation}\label{eq:Before}
\frac{-W|P|j}{j\cdot P + |\mathcal{C}|\sum_{\ell = 1}^j \ell}
\end{equation}
more over, in the path $\pi_{j+1} = \pi_j \pi^{j+1}$ the average weight in dimension $i$, for any prefix longer than $|\pi_j|$ is at least
\begin{equation}\label{eq:After}
\frac{-W|P|(j+1) -W(j+1)|\mathcal{C}|}{j\cdot P + |\mathcal{C}|\sum_{\ell = 1}^j \ell}
\end{equation}
Since the value of the sum $\sum_{\ell = 1}^j \ell$ is at least $\frac{j^2}{4}$, we get that for a large enough number $j_\epsilon$, for any prefix of $\pi$ that is longer than $j_\epsilon$, the average weight in dimension $i$ is at least $-\epsilon$, and the claim follows.

Thus $\LimInfAvgAutomat{A_i}(\pi)\geq r_i$ for every dimension $i$, which concludes the proof of Lemma~\ref{lem:ConnectionBetweenMultiCycleAndEmptiness} \pfbox
\end{proof}
\subsection{Proofs for subsection~\ref{subsect:LimInfAndSup}} 
\subsubsection{Proof of Lemma~\ref{prop:SupInfAvgDivideAndConcore}}
In this proof, w.l.o.g, we assume that both threshold vectors are the zero vector (that is, the vector $\VEC{0}$).

The proof for the direction from left to right is trivial.
To prove the converse direction, we denote by
 $\Automat{A}$ the $2k$-dimensional product automaton of the automata $A_1,\dots,A_k$, $B_1,\dots,B_k$;
recall that an infinite word corresponds to an infinite path in $\Automat{A}$;
hence we can assume that there exist infinite paths $\pi_1,\dots,\pi_k$ such that for every $j\in\RangeSet{1}{k}$:
\begin{quote}
$\LimInfAvgAutomat{A_i}(\pi_j) \geq 0$ for all $i\in\RangeSet{1}{k}$ ;
and
$\LimSupAvgAutomat{B_j}(\pi_j) \geq 0$
\end{quote}
Informally, we shall construct the witness path $\pi$, for which $\LimInfAvgAutomat{A_i}(\pi) \geq 0$ and $\LimInfAvgAutomat{B_i}(\pi) \geq 0$ for all $i\in\RangeSet{1}{k}$,
by following the path $\pi_j$ until the average weights in the corresponding dimensions of $A_1,\dots,A_k$ and $B_j$ are sufficiently close to $0$, and repeat the process for $1 + j\mod k$, and so on.

Formally:
We denote the value of the minimal weight that occur in the graph by $-W$.
For every $j\in\RangeSet{1}{k}$ and $\epsilon > 0$ we denote by $N_j^\epsilon$ the first position in the path $\pi_j$ such that in every position of $\pi_j$, that is greater than $N_j^\epsilon$, the average weight in all the dimensions that corresponds to $A_1,\dots,A_k$ is at least $-\epsilon$;
the reader should note that by definition, such $N_j^\epsilon$ always exists;
we denote $N^\epsilon = \max_{j\in\RangeSet{1}{k}} N_j^\epsilon$.
For every $m\in\Nat$, $\epsilon > 0$ and a finite path $\lambda$ we denote by $\pi_j^{m,\epsilon,\lambda}$ the shortest prefix of the path $\lambda \cdot \pi_j$, of length at least $m$, such that average weights in the prefix, in all the dimensions that correspond to $A_1,\dots,A_k$ and $B_j$, are at least $-\epsilon$ (note that by the definition of $\pi_j$, such prefix exists).
The following remark demonstrate the key property of the definitions above.
\begin{rem}\label{rem:keyPropOfDefAbove}
Let $\lambda$ be a finite path, and let $M_\epsilon = \frac{(W N^\epsilon)}{\epsilon}$.
For every $i,j\in\RangeSet{1}{k}$ and $\epsilon > 0$, the infinite path $\pi_i^{M_\epsilon,\epsilon,\lambda} \pi_j$ satisfies
\begin{quote}
In every position of the path, greater than $\pi_i^{M_\epsilon,\epsilon,\lambda}$, the average weight in the dimensions that correspond to $A_1,\dots,A_k$ is at least $-2\epsilon$.
\end{quote}
\end{rem}
We define an infinite sequence of finite paths $\lambda_0, \lambda_1, \lambda_2, \dots$, in the following way:
\begin{itemize}
\item $\lambda_0$ is an arbitrary path, for example: the first edge of $\pi_1$.
\item For $i>0$: we denote $\lambda_i = \lambda_{i-1} \cdot \pi_i^{M_{\epsilon_i},\epsilon_i,\lambda_{i-1}}$,
where $\epsilon_i = \frac{1}{i}$ and for $i > k$: $\pi_i \equiv \pi_{1 + (i\pmod k)}$.
\end{itemize}
We define $\pi$ to be the limit path of the sequence $\Set{\lambda_i}_{i=1}^\infty$;
by the construction of $\pi$ it follows that for every $\epsilon > 0$, in infinite many positions, the average weight of the dimension that corresponds to $B_i$ is at least $-\epsilon$, and that as of certain position, the average weight in all the dimensions that correspond to $A_1,\dots,A_k$ is at least $-\epsilon$.
Hence, by definition, for all $i\in\RangeSet{1}{k}$:
\begin{quote}
$\LimInfAvgAutomat{A_i}(\pi) \geq 0$ and $\LimSupAvgAutomat{B_i}(\pi) \geq 0$
\end{quote}
as required, and the proof of the lemma follows. \pfbox
\subsubsection{Proof of Lemma~\ref{prop:ConjSupAndInfBecomesInf}}
\begin{proof}
The direction from right to left is trivial;
to prove the converse direction, let us assume that the intersection $\InfAvgLan{B_1}{r^b_1} \cap (\bigcap_{i=1}^k \InfAvgLan{A_i}{r^a_i})$ is empty, and we shall prove that the intersection $\SupAvgLan{B_1}{r^b_1} \cap (\bigcap_{i=1}^k \InfAvgLan{A_i}{r^a_i})$ is also empty.
W.l.o.g we assume that $\VEC{r^a} = \VEC{r^b} = \VEC{0}$.
Let $\Automat{A}$ be the $k+1$-dimensional product automaton of the automata $A_1,\dots,A_k$ and $B_1$.
Let $\pi$ be an infinite path in $\Automat{A}$, we will show that either $\LimInfAvgAutomat{A_i}(\pi) < 0$ for some $i\in\RangeSet{1}{k}$ or that $\LimSupAvgAutomat{B_1}(\pi) < 0$.
We first observe that the automaton $\Automat{A}$ does not have a $\VEC{0}$ multi-cycle; 
this fact follows from the assumption that the intersection $\InfAvgLan{B_1}{0} \cap (\bigcap_{i=1}^k \InfAvgLan{A_i}{0})$ is empty and from Lemma~\ref{lem:ConnectionBetweenMultiCycleAndEmptiness}.
Hence, by Lemma~\ref{prop:NoMultiCycleImpliesNegPayoff}, there exist constants $m\in\Nat$ and $c > 0$ such that for every prefix $\pi '$ of the path $\pi$ there exists a dimension $i$ for which the weight of $\pi '$ in dimension $i$ is at most $m - c|\pi '|$;
let $j$ be the corresponding dimension of the automaton $B_1$, and let $\pi^*$ be the longest prefix with average weight at most $-\frac{c}{2}$ in dimension $j$ (note that $\pi^*$ does not necessarily exist).
We consider two disjoint cases:
In the first case, there exists such $\pi^*$, and then, by definition $\LimSupAvgAutomat{B_1}(\pi) \leq -\frac{c}{2} < 0$ and the claim follows.
In the second case, the path $\pi$ has infinitely many prefixes with average weight at least $-\frac{c}{2}$ in dimension $j$;
therefore (by Lemma~\ref{prop:NoMultiCycleImpliesNegPayoff}) there must exist a dimension $i$ for which the average weight in dimension $i$ is at most $-\frac{c}{2}$ for infinitely many prefixes of $\pi$;
hence, $\LimInfAvgAutomat{A_i}(\pi) \leq -\frac{c}{2} < 0$, and the claim follows.

To conclude, if the intersection $\InfAvgLan{B_1}{0} \cap (\bigcap_{i=1}^k \InfAvgLan{A_i}{0})$ is empty, then
for every infinite path $\pi$, either $\LimSupAvgAutomat{B_1}(\pi) < 0$ or there exists some $i\in\RangeSet{1}{k}$ such that $\LimInfAvgAutomat{A_i}(\pi) < 0$,
which concludes the proof of Lemma~\ref{prop:ConjSupAndInfBecomesInf}. \pfbox
\end{proof}
\subsection{Proofs for subsection~\ref{subsect:maxFreePSPACE}} 
\subsubsection{Proof of Lemma~\ref{prop:PropsOfMaxFreeConstrains}}
\begin{proof}
To prove the first item of Lemma~\ref{prop:PropsOfMaxFreeConstrains} we present a detailed description of the $\MAX$-free constraints for an expression $E$ and a threshold $\nu$.
Let $\Automat{A}$ be the $k$-dimensional product automaton of the automata that occur in the expression $E$; w.l.o.g we assume that for some $j$, the first $j$ dimensions of $\Automat{A}$ correspond to the $\LimInfAvg$ automata $A_1,\dots,A_j$, and the last $k-j$ dimensions correspond to the $\LimSupAvg$ automata $A_{j+1},\dots,A_k$.
Let $\VEC{r}$ be a $2k$-dimensional vector of variables;
let $\mathbb{C}$ be the set of all simple cycles in $\Automat{A}$; and we define a variable $X^i_c$ for every cycle $c$ in $\mathbb{C}$ and a dimension $i\in\RangeSet{j+1}{k}$;
then the $\MIN$-only constraints for the expression $E$ and the threshold vector $(r_1,\dots,r_k)$ are as follows:
\begin{equation}\label{eq:MinOnlyConstrains}
\sum_{c\in \mathbb{C}} X^i_c \WtFunc_m(c) \geq r_m \mbox{ for every $i\in\Set{j+1,\dots,k}, m\in\Set{1,\dots,j,i}$}
\end{equation}
\begin{equation}
\sum_{c\in \mathbb{C}} |c|X^i_c =  1 \mbox{ for every $i\in\Set{j+1,\dots,k}$}
\end{equation}
\begin{equation}
X_c^i \geq 0 \mbox{ for every $i\in\RangeSet{j+1}{k}$, and $c\in \mathbb{C}$}
\end{equation}
and the $\MAX$-free constraints are the $\MIN$-only constraints along with the constraints
\begin{equation}
M_{E} \times \VEC{r} \geq \VEC{b_\nu}
\end{equation}
where $M_{E}$ and $\VEC{b_\nu}$ are the $2k \times 2k$ matrix and the $2k$-dimensional vector from Lemma~\ref{prop:ReductionFromMaxFreeToConjunction}.

The proof of Lemma~\ref{prop:PropsOfMaxFreeConstrains}(1) follows immediately from the definition of the $\MAX$-free constraints.

To prove the additional two items of the lemma, we observe that the $\MAX$-free constraints for the expression $E$ and the threshold $\nu$ remain linear  when the threshold $\nu$ is a variable;
hence, the problem of computing the maximum threshold $\nu$ for which the $\MAX$-free constraints are feasible amounts to the following linear-programming problem:
\begin{quote}
Find the maximum $\nu$ subject to the $\MAX$-free constraints for expression $E$ and threshold $\nu$.
\end{quote}
Let $W$ be a bound on the absolute value of the weights of $\Automat{A}$;
clearly the $\MAX$-free constraints are feasible for the threshold $-W$, and are infeasible for the threshold $+2W$;
hence, the domain of feasible thresholds is bounded and nonempty, and by standard properties of linear programming~\cite{LinearProgramming}, it follows that a maximum threshold exists.
Moreover, we observe that every coefficient in the $\MAX$-free constraints can be encoded by polynomial number of bits (as the length of a simple cycle in the product automaton is at most exponential in the size of the input),
and that the number of constraints, which are not of the form of $x\geq 0$, is $O(k^2)$;
we denote by $D$ the maximal coefficient in the $\MAX$-free constraints, by standard properties of linear programming~\cite{LinearProgramming}, it follows that the maximum value of $\nu$ is obtained when at most $O(k^2)$ variables are assigned with rational nonzero values, and the result of the linear-programming (that is, the maximum threshold) is a rational with numerator and denominator bounded by $D^{O(k^2)}$;
therefore only $O(k^2)\log(D)$ bits are required to encode the maximum threshold.

To conclude, we proved that for the maximum threshold $\nu^*$ for which the $\MAX$-free constraints are feasible, there is a solution for the corresponding $\MAX$-free constraints such that at most $O(k^2)$ variables are assigned with nonzero values; and the threshold $\nu^*$ requires only polynomial number of bits to encode;
hence the last two items of Lemma~\ref{prop:PropsOfMaxFreeConstrains} immediately follows. \pfbox
\end{proof}

\subsubsection{Proof of Lemma~\ref{lem:MaxFreeEmptInPSPACE}}
\begin{proof}
Recall that we can decide the non-emptiness of a $\MAX$-free expression by checking the feasibility of its corresponding $\MAX$-free constraints,
and that a rational solution for the $\MAX$-free constraints corresponds to vectors of thresholds and a set of multi-cycles, each multi-cycle with an average weight that matches its corresponding threshold vector;
by Lemma~\ref{prop:PropsOfMaxFreeConstrains}(2) the number of different simple cycles that occur in the witness multi-cycles set is at most $t$.
We also observe that if a multi-set of cycles (that are not necessarily simple) with average weight vector $\VEC{\nu}$ exists, then a $\VEC{\nu}$ multi-cycle (of simple cycles) also exists, since we can decompose every non-simple cycles to a set of simple cycles;
thus, a $\VEC{\nu}$ multi-cycle exists iff there exists a multi-set of \emph{short} cycles, where the length of each cycle in the multi-set is at most the number of vertices in the graph (note that in particular, every simple cycle is short).

Hence, we can decide the feasibility of the $\MAX$-free constraints in the following way:
First, we guess $t$ weight vectors of $t$ short cycles that occur in the same SCC of the product automaton of all the automata that occur in the expression;
second, we construct the $O(k^2)$ constraints of the $\MAX$-free constraints and assign zero values to all the variables of the non-chosen cycles;
third, we check the feasibility of the formed $O(k^2)$ constraints, where each constraint has at most $t + 1$ variables.

Note that we can easily perform the last two steps in polynomial time (as the values in the weight vector of every short cycle can be encoded by polynomial number of bits);
hence, to prove the existence of a PSPACE algorithm, it is enough to show
how to encode (and verify by a polynomial-space machine) $t$ average weight vectors of $t$ short cycles that belong to one SCC of the product automaton.

The encoding technique is straight forward and standard;
we present its details only for the purpose of self-containment.
We based the encoding on the observations that a vertex in the product automaton is a $k$-tuple of states, and that a short cycle is characterized by:
(a)~its length, which is at most the size of the product automaton (that is, at most exponential in the size of the expression); (b)~its initial (and final) vertex (that is, a $k$-tuple of states); and (c)~a sequence of alphabet symbols that corresponds to the path from its initial vertex to its end vertex.

Thus, we encode $t$ average weights by the sequence 
\begin{quote}
$(w_1,\dots,w_t)\#(a_1,b_1,c_1)\#(a_2,b_2,c_2)\#\dots \#(a_t,b_t,c_t)\#\pi_{1\to 2\to \dots \to t\to 1}$
\end{quote}
where $w_1,\dots,w_t$ are the $t$ average weights;
$a_i$ ,$b_i$ and $c_i$ are respectively the length, initial vertex and the path (that is, sequence of alphabet symbols) of the cycle with the weight $w_i$;
and $\pi_{1\to 2\to \dots \to t\to 1}$ is a sequence of symbols that corresponds to a cyclic path between all the initial vertices of all $t$ cycles.

To verify the encoding, we simply simulate, in parallel, each of the $k$ weighted automata, and store the weight vectors of each short cycle, which requires at most polynomial number of bits\footnote{Note that in general, the sum of exponential number of rationals may require exponential number of bits; however, in our case, there are only $|\Automat{A}|$ different rational numbers, hence their lowest common denominator has only polynomial number of bits, and their sum can be encoded by polynomial number of bits.};
we verify that all the cycles are in the same SCC, given the witness $\pi_{1\to 2\to \dots \to t\to 1}$, in similar way;
finally, all that is left is to verify that the vector $(w_1,\dots,w_t)$ corresponds to the real average weights of the cycles;
in addition, we reject the witness if its number of bits exceeds the  (at most exponential) length threshold of $t^3 |\Automat{A}|$, where $\Automat{A}$ is the product automaton of all the automata occur in the expression.

To conclude, we proved that there is a PSPACE algorithm that decides the feasibly of the $\MAX$-free constraints, and Lemma~\ref{lem:MaxFreeEmptInPSPACE} follows. \pfbox
\end{proof}

\subsubsection{Proof of Lemma~\ref{lem:MaxFreeMaxInPSPACE}}
\begin{proof}
By definition, there exists an infinite word $w$ such that $E(w) \geq \nu$ iff the expression $E$ is nonempty with respect to threshold $\nu$.
By Lemma~\ref{prop:PropsOfMaxFreeConstrains}(3) and by the equivalence of the feasibility of the $\MAX$-free constraints and the emptiness of the expression, the maximum threshold for which the expression $E$ is nonempty is a $t$-bits rational number (and moreover such maximum threshold exists);
hence we can find this maximum threshold by checking the emptiness of $E$ for all thresholds that are $t$-bits rational numbers, and return the maximal such threshold (possible optimization is to do binary search over all $t$-bits thresholds);
this can be done in polynomial space due to Lemma~\ref{lem:MaxFreeEmptInPSPACE}.
\pfbox  
\end{proof}

\section{Proofs for Section~\ref{sect:PSPACEforAllProblems}}
\subsection{Proof of Theorem~\ref{lem:MaxForExpression}}
\begin{proof}
First, we claim that if the number of $\MAX$ operators in the expression $E$ is $m > 0$, then we can construct in linear time two expressions $E_1$ and $E_2$, each with at most $m-1$ $\MAX$ operators and of size at most $|E|$, such that $E = \MAX(E_1,E_2)$.
We prove the claim by induction on the number of (any) operators in the expression:
if the number of operators in the expression is zero, the claim is trivially satisfied;
otherwise let $E = \OP(F,G)$ such that the expression $F$ has at least one $\MAX$ operator;
as $F$ has strictly fewer operators (as compared to $E$), by the inductive hypothesis there exist two expressions $F_1$ and $F_2$, each with at most $m-1$ $\MAX$ operators, and of length at most $|F|$, such that $F = \MAX(F_1,F_2)$;
the reader can verify that $E_1 = \OP(F_1,G)$ and $E_2 = \OP(F_2,G)$ satisfy the claim, that is, $E=\MAX(E_1,E_2)$ and $|E_1|,|E_2|\leq |E|$.

Second, we observe the fact that if $E = \MAX(E_1,E_2)$,
then
\begin{quote}
$\sup_{w\in\Sigma^\omega} E(w) =
			\max(\sup_{w\in\Sigma^\omega} E_1(w),
				  \sup_{w\in\Sigma^\omega} E_2(w))$
\end{quote}
Hence, in order to compute the maximum value of $E$, we simply compute the maximum value of $E_1$, the maximum value of $E_2$ and return the maximum of the two values.
If $E_1$ and $E_2$ are $\MAX$-free expressions, then thanks to Lemma~\ref{lem:MaxFreeMaxInPSPACE} we can compute their maximum values by polynomial space Turing machine;
in the general case, if the number of $\MAX$ operators in the expression $E$ is $m$, then we construct in linear time two expressions $E_1$ and $E_2$, each with at most $m-1$ $\MAX$ operators, such that $E = \MAX(E_1,E_2)$; then we recursively compute the maximum value of $E_1$ and the maximum value of $E_2$ and return the maximum of these values.

The reader can easily verify that the procedure we described requires only polynomial space to run, and thus the proof of Theorem~\ref{lem:MaxForExpression} follows. \pfbox
\end{proof}

\subsection{Lower bounds}
In this subsection we will establish PSPACE lower bounds for the emptiness, universality, language inclusion and language equivalence problems;
we obtain the PSPACE lower bound for the emptiness problem by a reduction from the emptiness problem for intersection of regular languages (which is PSPACE-hard~\cite{IntersectionOfRegularLanguagesIsPSPACEComplete});
for the universality problem we obtain the bound by a reduction from the universality problem for union of regular languages (which is the complement of the first problem, and hence, also PSPACE-hard);
and the lower bounds for the language inclusion and equivalence problems are obtained by a reduction from the universality problem (of mean-payoff expressions).

To present the reduction from the emptiness problem for intersection of regular languages we define, for a language of \textbf{finite} words $L\subseteq \Sigma^*$, and for a symbol $\xi\notin \Sigma$, the following function, for every infinite word $w\in(\Sigma\cup\Set{\xi})^\omega$:
\begin{quote}
\[ f^L(w) = \left\{ \begin{array}{ll}
        +1 & \mbox{if $w\in L\cdot\xi \cdot(\Sigma + \xi)^\omega$}\\
        -1 & \mbox{otherwise}\end{array} \right. \]
\end{quote}
It is easy to verify that if the language $L$ is recognizable by a finite-state automaton $A$, then we can construct in linear time a weighted automaton $\Automat{A}$ such that $\LimInfAvgAutomat{\Automat{A}} \equiv f^{L(A)}$;
and that for any $k$ finite-state automata $A_1,\dots,A_k$ over the alphabet $\Sigma$, the intersection $\bigcap_{i=1}^k L(A_i)$ is nonempty iff the expression $E = \MIN(\LimInfAvgAutomat{\Automat{A}_1},\dots,\LimInfAvgAutomat{\Automat{A}_k})$ is nonempty with respect to threshold $0$ and alphabet $\Sigma\cup\Set{\xi}$;
thus, the emptiness problem for mean-payoff expressions is PSPACE-hard.

We prove the PSPACE hardness of the universality problem in a similar way;
we define the next function for every infinite word $w\in(\Sigma\cup\Set{\xi})^\omega$:
\begin{quote}
\[ g(w) = \left\{ \begin{array}{ll}
        +1 & \mbox{if $w\in\Sigma^\omega$}\\
        -1 & \mbox{otherwise}\end{array} \right. \]
\end{quote}
we denote by $G$ the (minimal) weighted automaton for which $\LimInfAvgAutomat{G}(w) \equiv g$ (surely, such automaton exists);
and we observe that the union of the regular languages $\bigcup_{i=1}^k L(A_i)$ is universal iff the expression $E = \MAX(\LimInfAvgAutomat{\Automat{A}_1},\dots,\LimInfAvgAutomat{\Automat{A}_k},\LimInfAvgAutomat{G})$
is universal with respect to threshold $0$ and alphabet $\Sigma\cup\Set{\xi}$ (recall that $\Automat{A}_i \equiv f^{L(A_i)})$.
Thus, the universality problem for mean-payoff expressions is PSPACE-hard.

The reductions from the universality problem to the language inclusion and equivalence problems are trivial;
let $Z$ denote a weighted automaton for which $\LimInfAvgAutomat{Z}(w) \equiv 0$ (for example an automaton where all the weights of the edges are zero,
and let $E_0$ denote $\LimInfAvgAutomat{Z}$;
then the expression $E$ is universal with respect to threshold $0$ iff $E \geq E_0$,
and iff the expression $\MIN(E,E_0)$ is equivalent to the expression $E_0$.
Since the universality problem is PSPACE-hard even for threshold $0$,
 the next lemma follows.
\begin{lem}\label{lem:LowerBounds}
For the class of mean-payoff automaton expressions, the quantitative
emptiness, universality, language inclusion, and equivalence problems are PSPACE-hard.
\end{lem}
}

\end{document}